\documentclass[aps,twocolumn,floatfix,showpacs,showkeys]{revtex4-1}
\usepackage{graphicx}
\usepackage{natbib}
\usepackage{epstopdf}
\usepackage{amsmath}
\usepackage{amssymb}
\usepackage{mathbbol}
\usepackage{braket}
\usepackage[hyperindex,colorlinks,urlcolor=blue]{hyperref}

\usepackage{xcolor}

\begin{document}

\title{Quantum phase transitions in the driven dissipative Jaynes-Cummings oscillator: from the dispersive regime to resonance}

\author{Th. K. Mavrogordatos}\email{t.mavrogordatos@ucl.ac.uk}
\affiliation{Department of Physics and Astronomy, University College London,
Gower Street, London, WC1E 6BT, United Kingdom}

\date{\today}

\begin{abstract}
We follow the passage from complex amplitude bistability to phase bistability in the driven dissipative Jaynes-Cummings oscillator. {\it Quasi}distribution functions in the steady state are employed, for varying qubit-cavity detuning and drive parameters, in order to track a first-order dissipative quantum phase transition up to the critical point marking a second-order transition and spontaneous symmetry breaking. We demonstrate the photon blockade breakdown in the dispersive regime, and find that the coexistence of cavity states in the regime of quantum bistability is accompanied by pronounced qubit-cavity entanglement. Focusing on the r\^{o}le of quantum fluctuations in the response of both coupled quantum degrees of freedom (cavity and qubit), we move from a region of minimal entanglement in the dispersive regime, where we derive analytical perturbative results, to the threshold behaviour of spontaneous dressed-state polarization at resonance.
\end{abstract}

\maketitle
\section{Introduction}

The Jaynes-Cummings (JC) oscillator is an archetypal source of intricate quantum nonlinear dynamics arising from the coupling of a quantized electromagnetic mode inside a resonator (cavity) to a two-level system (qubit) \cite{JaynesCummings}. The behaviour of quantum nonlinear oscillators has been a subject of intense theoretical investigation (for an overview see Chapter 7 of \cite{DykmanBook}) providing at the same time the basis for numerous experiments in cavity and circuit quantum electrodynamics (see for example \cite{BishopNL} where the extended JC oscillator is driven out of equilibrium in the presence of dissipation). In addition, controlled light-matter interaction has shifted the center of interest in phase transitions from condensed matter to quantum optics. 

Amongst the most discussed light-matter quantum phase transitions in the literature are the Dicke phase transition \cite{Dicke1, Dicke2}, which is explicitly dissipative (with cooperative resonance fluorescence as its driven variant \cite{CoopFl}), and the laser which exhibits a second-order phase transition out of equilibrium \cite{laser}. In comparison to those, however, the driven JC model is fundamentally different as it deals with the interaction of one field mode with one quantum object, the qubit, necessitating the reappraisal of the r\^{o}le of quantum fluctuations and a different definition of the thermodynamic limit \cite{CarmichaelBook1, CarmichaelBook2, PhotonBlockade, Domokos}. 
\begin{figure*}
\centering 
\includegraphics[width=4.5in]{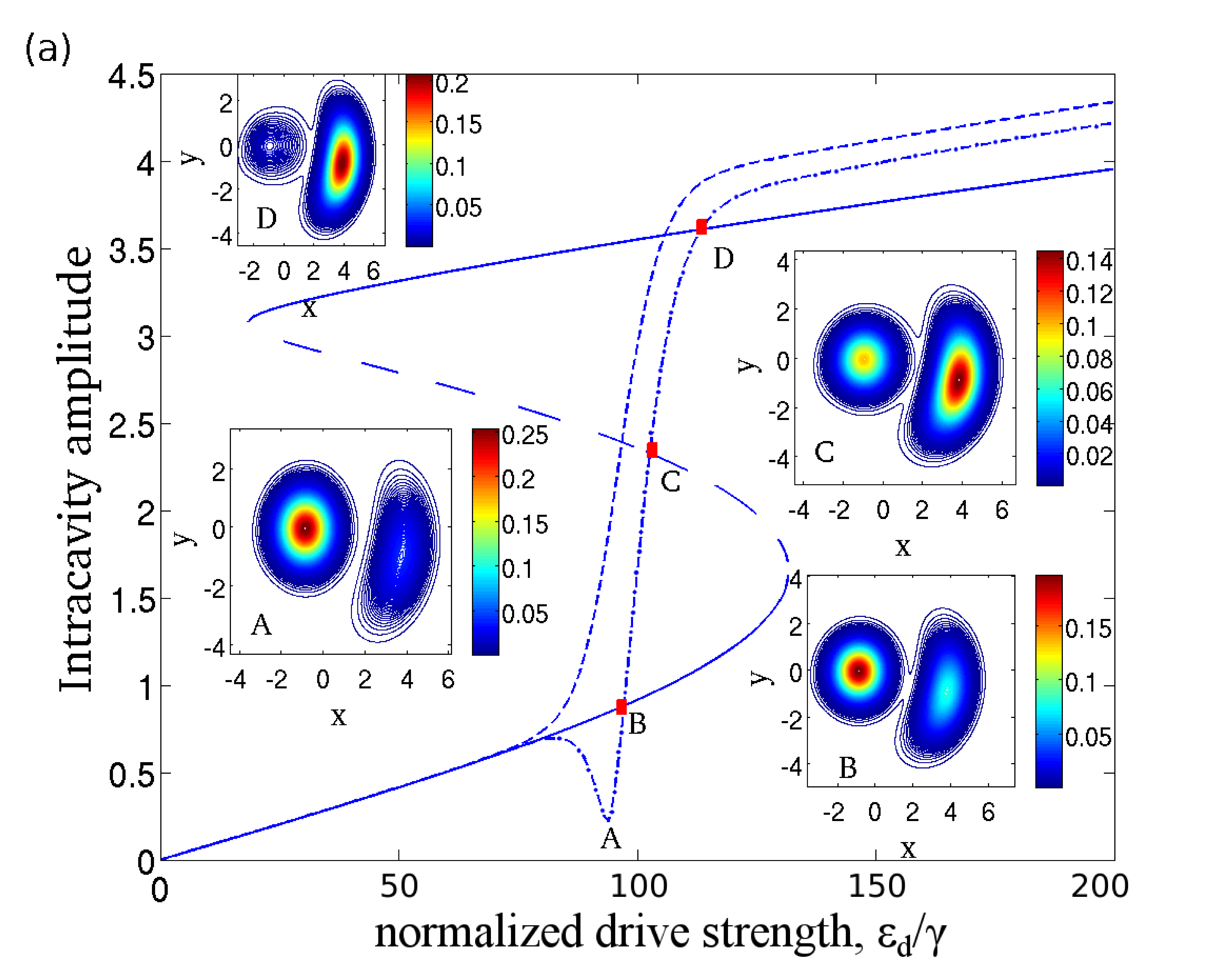}
\includegraphics[width=2.4in]{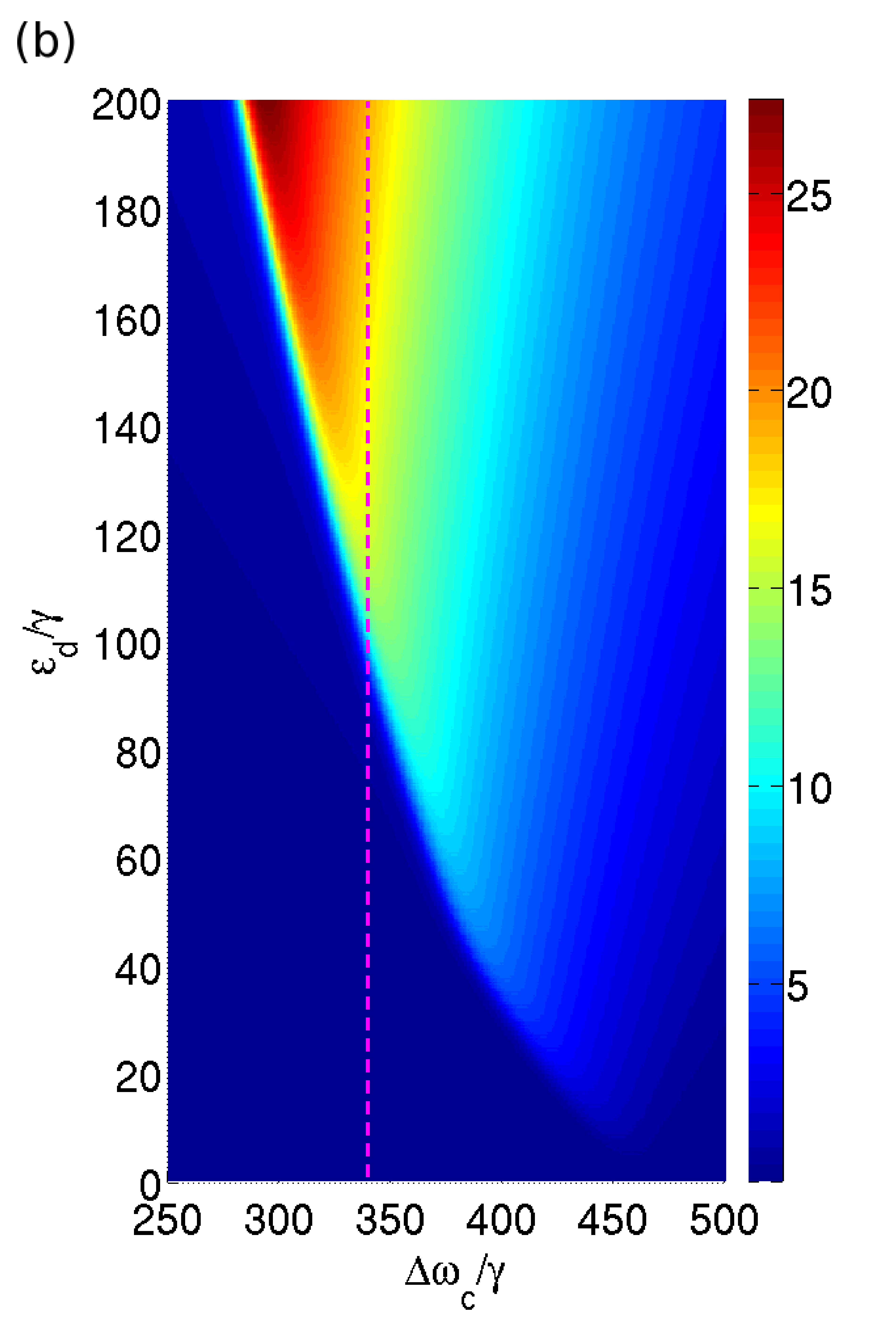}
\caption{{\it Dispersive amplitude bistability.} \textbf{(a)} Steady-state intracavity amplitude as a function of the normalized drive strength $\varepsilon_d/\gamma$ in the semiclassical and the quantum description for $\Delta\omega_c/\gamma=340$. The semiclassical bistability curve (solid line, with the sparsely dashed part corresponding to the unstable branch) depicting $|\alpha|$ is superimposed on top of the quantum amplitude curves $\sqrt{\braket{a^{\dagger}a}}$ (thinly dashed line) and $\left|\braket{a}\right|$ (dashed-dotted line). The latter exhibits the characteristic coherent cancellation dip (point A) of the Duffing oscillator, and intersects the semiclassical bistability curve in three points (B,C,D). For the marked points A, B, C, D we plot the {\it quasi}distribution function $Q(x + iy)$ for the corresponding intracavity field amplitude. \textbf{(b)} Average cavity photon number $\braket{a^{\dagger}a}$ as a function of the drive parameters. The dashed line indicates the driving frequency selected for \textbf{(a)}. Parameters: $g/\delta=0.14$, $2\kappa/\gamma=12$, $g/\gamma=3347$, $n_{\rm scale}=12.68$.}
\label{fig:IntAmp} 
\end{figure*}
The open coherently-driven dissipative qubit-cavity system yields a bistable response where quantum fluctuations are responsible for switching between two metastable states that are long-lived in relation to the characteristic cavity and qubit decay times \cite{Savage}. The $\sqrt{n}$ splitting of the JC energy levels is a unique feature determining the nature of bistability both at resonance, where the cavity and qubit bare frequencies coincide, and in the dispersive regime, where the cavity and qubit are strongly detuned in relation to their dipole coupling strength \cite{Fink, PhotonBlockade, DispersiveTransform}. 

At resonance, the mean-field nonlinearity diverges for zero photon number, while in the dispersive regime bistability builds up in a perturbative fashion with no associated threshold, unlike the laser. The perturbative approach becomes inadequate when the qubit participates actively in the bistable switching for stronger driving \cite{simultaneousbistability}. In that regime, the system response comprises an average over spontaneous switching between the metastable mean-field steady states where both the cavity and qubit are significantly excited. For complex amplitude bistability switching occurs between a dim (with lower $n$) and a bright (with higher $n$) state, while in phase bistability both states have the same magnitude and opposite phases following a transition from a discrete to a continuous spectrum in the system {\it quasi}energies \cite{WallsBook, PhotonBlockade, CarmichaelBook2}. In first-order dissipative phase transitions for the cavity field, weaker coupling implies a bigger photon number required for the nonlinearity to manifest itself, yielding a response which is a non-analytic function of the drive \cite{DykmanBook, PhotonBlockade}.

Motivated by the current experimental and theoretical interest in the response of quantum nonlinear oscillators, in this letter we track amplitude bistability, from its origin in the dispersive regime, up to a critical point at resonance, where phase bistability takes over. Mean-field results guide us to extract the relevant scaling parameters used to define the ``thermodynamic limit'' for this driven resonator in which the number of photons is not conserved. We present contour plots of {\it quasi}distribution functions for the cavity field, showing the passage from amplitude to phase bistability, and invoke the entanglement entropy, calculated via the reduced qubit density matrix, in order to demonstrate the active participation of both quantum degrees of freedom in the emerging bimodality. We show further that, closer to resonance, enhanced multi-photon transitions appear for weak cavity excitation, followed by a breakdown of the photon blockade with stronger driving \cite{Domokos, PhotonBlockade}.

\section{The dispersive JC model}

The JC Hamiltonian describes the interaction between a single resonant cavity mode and a qubit; however, it does not account directly for the coupling to the environment which is included only in the formulation of the Master Equation (ME) \cite{SpontaneousDressedState}. After adding dissipation in a frame rotating with the frequency $\omega_d$ of the coherent driving field, the Lindblad ME for the reduced system density operator $\rho$ (for a system Hamiltonian in the rotating wave approximation and setting $\hbar=1$ for convenience) reads \cite{DuffingWalls, CarmichaelBook2}
\begin{equation}
\begin{split}
\dot{\rho}& =i \Delta\omega_c [a^{\dagger}a, \rho] + i \Delta\omega_q [\sigma_{+}\sigma_{-}, \rho]+ g [a^\dagger \sigma_{-} - a\sigma_{+}, \rho]\\
&+ [\varepsilon_d a^{\dagger} -\varepsilon_d^{*}a, \rho]+ \kappa(2a\rho a^{\dagger}-\rho a^{\dagger}a - a^{\dagger}a \rho)\\
& + (\gamma/2)(2\sigma_{-} \rho \sigma_{+}-\rho \sigma_{+}\sigma_{-} - \sigma_{+}\sigma_{-}\rho).
\end{split}
\label{ME}
\end{equation}
\begin{figure*}
\centering
\includegraphics[width=3.5in]{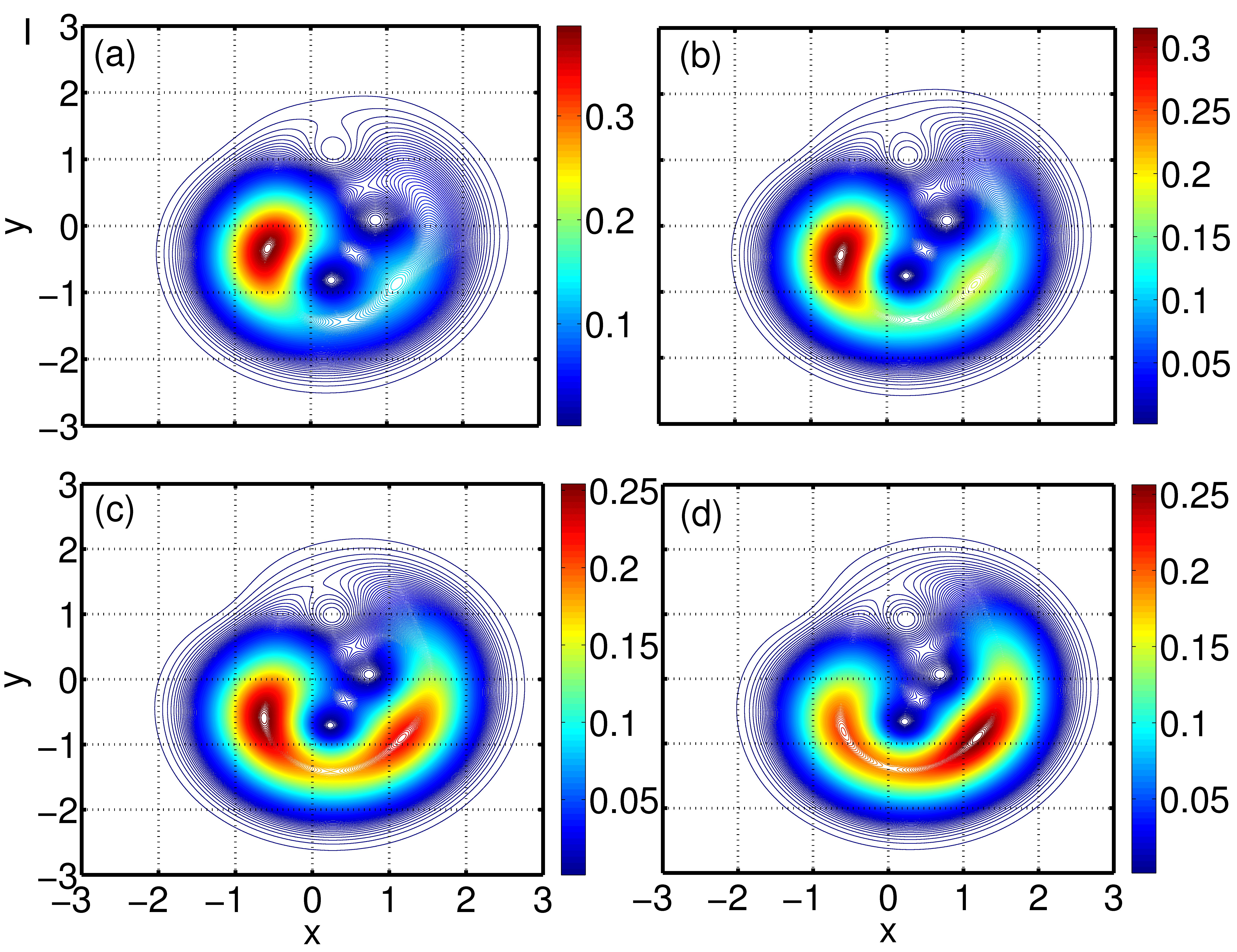}
\includegraphics[width=3.5in]{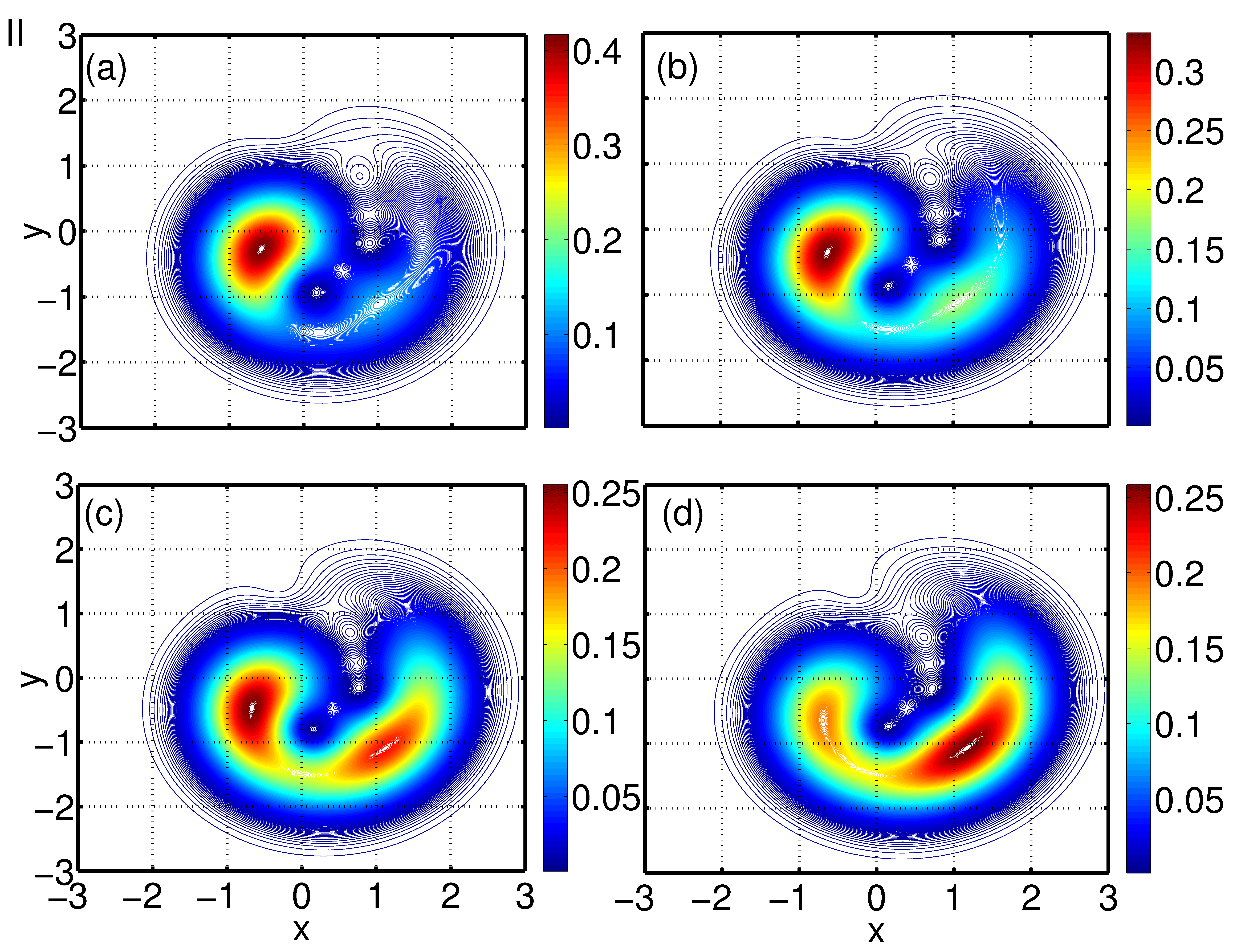}
\caption{{\it The effective Kerr nonlinearity.} Joint {\it quasi}distribution function $W(x + iy)$ for $\Delta \omega_c/\kappa=72.50$ and four different values of the drive strength: $\varepsilon_d/\kappa=2.17, 2.33, 2.50, 2.67$ in \textbf{(a)}-\textbf{(d)} respectively, using Eq. \eqref{WignerFinal} (\textbf{Panel I}) and the solution of Eq. \eqref{ME} for the reduced cavity density matrix (\textbf{Panel II}). Parameters: $g/\delta=0.14$, $2\kappa/\gamma=12$, $g/\gamma=3347$, $n_{\rm scale}=12.68$.}
\label{fig:WDuff}
\end{figure*}
In the right hand side of the above equation, $\Delta\omega_{c,q}=\omega_d-\omega_{c,q}$ are the detunings of the cavity resonance frequency $\omega_c$ and the qubit bare frequency $\omega_q$ from the frequency of the drive, coupled to a resonant cavity mode with photon annihilation and creation operators $a$ and $a^{\dagger}$, respectively. The inversion operator $\sigma_z$ is related to the raising (lowering) operators $\sigma_{+}$ ($\sigma_{-}$) for the qubit with two states $\ket{{\rm g}}$ (ground), $\ket{{\rm e}}$ (excited) via $\sigma_z=2\sigma_{+}\sigma_{-}-1$. The cavity mode is dipole-coupled to the qubit with strength $g$, while the classical coherent field (with very high photon occupancy) is coupled to the resonant cavity mode with strength $\varepsilon_d$ (also called {\it drive amplitude}). The cavity field is also coupled to a Markovian thermal bath at zero temperature, which induces a photon loss rate of $2\kappa$. In addition to photon dissipation, there is also spontaneous emission to modes different than the resonant cavity mode, with rate $\gamma$. The {\it strongly dispersive regime} with weak spontaneous emission is defined through a qubit-cavity detuning such that $\delta \equiv |\omega_c -\omega_q| \gg g \gg 2\kappa \gg \gamma$. In our results we show complex amplitude bistability for the intracavity field in the following region of the drive phase space: $0 \leq\Delta\omega_c \leq g^2/\delta$ and $\varepsilon_d < g^2/\delta < g$ (with $\gamma/(2\kappa) \approx 0.1$). At resonance, where $\delta=0$, phase bistability is associated with a particular point in the phase space: $(\Delta \omega_c=0,\, \varepsilon_d=g/2)$, and a threshold behaviour. Throughout our work, we solve numerically the Lindblad ME. We perform numerical simulations (employing the open-source Matlab software \textit{Quantum Optics Toolbox}) in a truncated Hilbert space for an initial product state with the qubit in the ground state and the cavity in a Fock state with zero photons [i.e. the initial ground state $\rho(0)=(\ket{n=0}\bra{n=0}) \otimes (\ket{{\rm g}}\bra{{\rm g}})$]. The steady-state results are independent of the initial state, while convergence with respect to the number of photon levels has been ensured. The validity of the rotating wave approximation in the dispersive regime we are considering is checked against Fig. 1 of Ref. \cite{RWAval} and Fig. 1 of Ref. \cite{BishopJC}. The former shows that the counter-propagating terms can be omitted without affecting the physical picture even for the maximum cavity-qubit detuning considered, since the coupling strength $g$ remains sufficiently smaller than the bare cavity frequency ($g/\omega_c \approx 0.03$), while the latter demonstrates good agreement between theory and experiment in a similar parameter regime. 

In Fig. \ref{fig:IntAmp} [see both frames (a) and (b)] we depict the the intracavity photon field (in the coherent state space with $\ket{\alpha}\equiv \Ket{x+iy}$) within a drive region where the quantum fluctuations are responsible for the deviation from the mean-field predictions. For low driving strengths, $|\alpha|, \sqrt{\braket{a^{\dagger}a}}$ and $\left| \braket{a} \right|$ coincide and the cavity is in the dim state, resembling a vacuum state with Gaussian distribution. With increasing $\varepsilon_d/\kappa$, the bright state accrues probability resulting in the coherent cancellation we observe at the point A. As we follow the curve for $\left| \braket{a} \right|$, probability transfers from the dim to the bright state crossing the boundary of a first-order dissipative quantum phase transition. The complex steady-state semiclassical intracavity amplitude $\alpha$ obeys the equation \cite{WallsBook}
\begin{equation}\label{nSC}
\alpha=-\frac{i\varepsilon_d}{\tilde{\kappa}} \left[1 + \frac{2g^2(\kappa - i \Delta \omega_c)^{-1} (\gamma - 2i\Delta \omega_q)^{-1}}{1 + \displaystyle \frac{8g^2|\alpha|^2}{(\gamma^2 + 4\Delta\omega_q^2)}} \right]^{-1},
\end{equation} 
with $\tilde{\kappa}=\kappa - i \Delta \omega_c$, which predicts two metastable states (dim and bright) and one unstable state that vanishes in the presence of fluctuations. In contrast to the mean-field prediction, the curve depicting $\left| \braket{a} \right|$ does not exhibit any bistability. Quantum fluctuations out of equilibrium manifest themselves through the absence of a Maxwell construction, since the line $\left| \braket{a} \right|$ does not cut the semiclassical curve in two equal areas. Furthermore, the curve for $\sqrt{\braket{a^{\dagger}a}}$ does not exhibit the coherent cancellation dip, which is hence solely a quantum phase effect. These considerations hold also for the driven Duffing oscillator (the reader is referred to Fig. 1 of \cite{DuffingWalls}) as a consequence of non-constant diffusion coefficients in the corresponding Fokker-Planck equation; here, however, we cannot formulate such an equation due to the active participation of the qubit \cite{CarmichaelBook1, CarmichaelBook2}. 
\begin{figure*}
\centering
\includegraphics[width=3.5in]{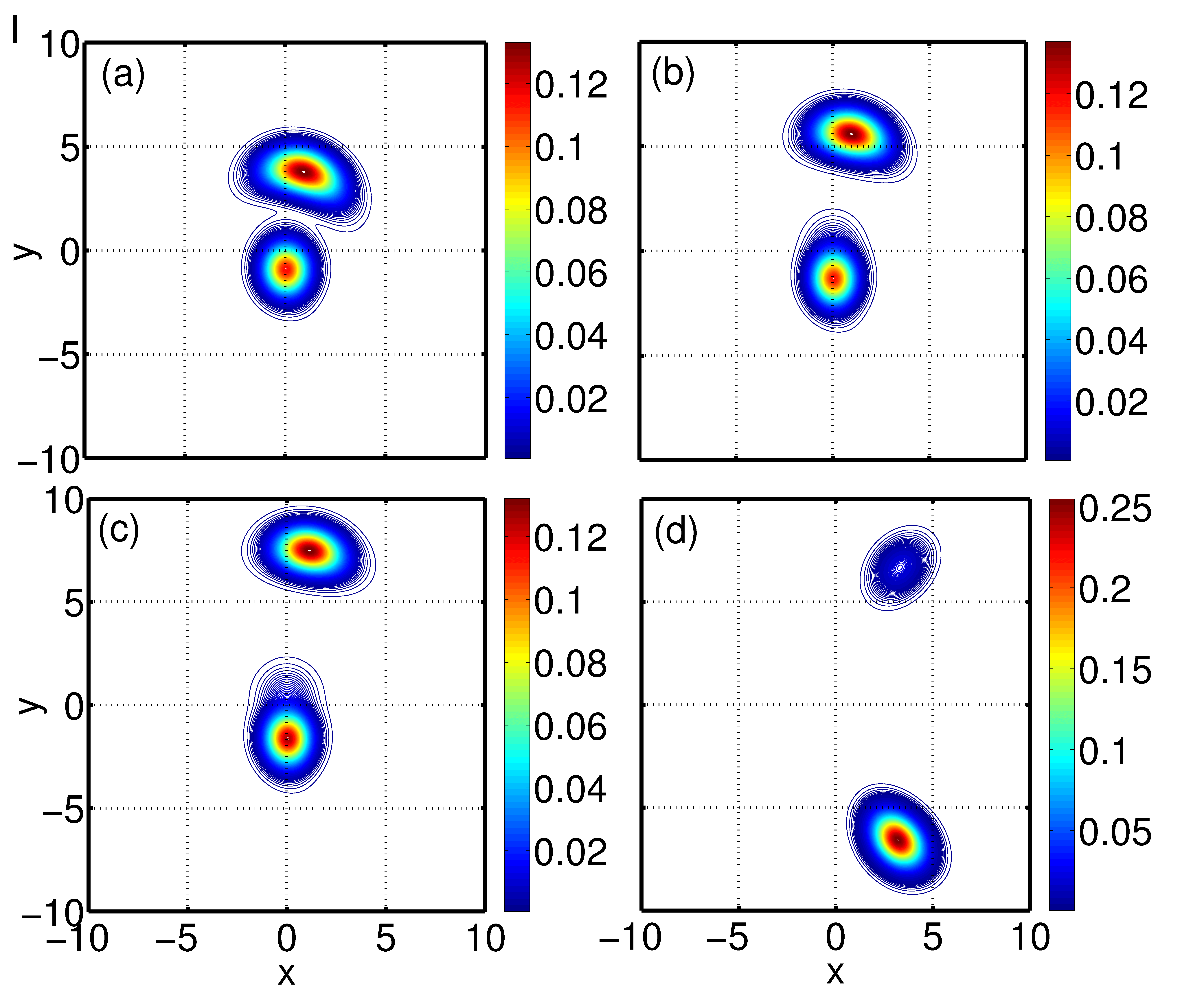}
\includegraphics[width=3.5in]{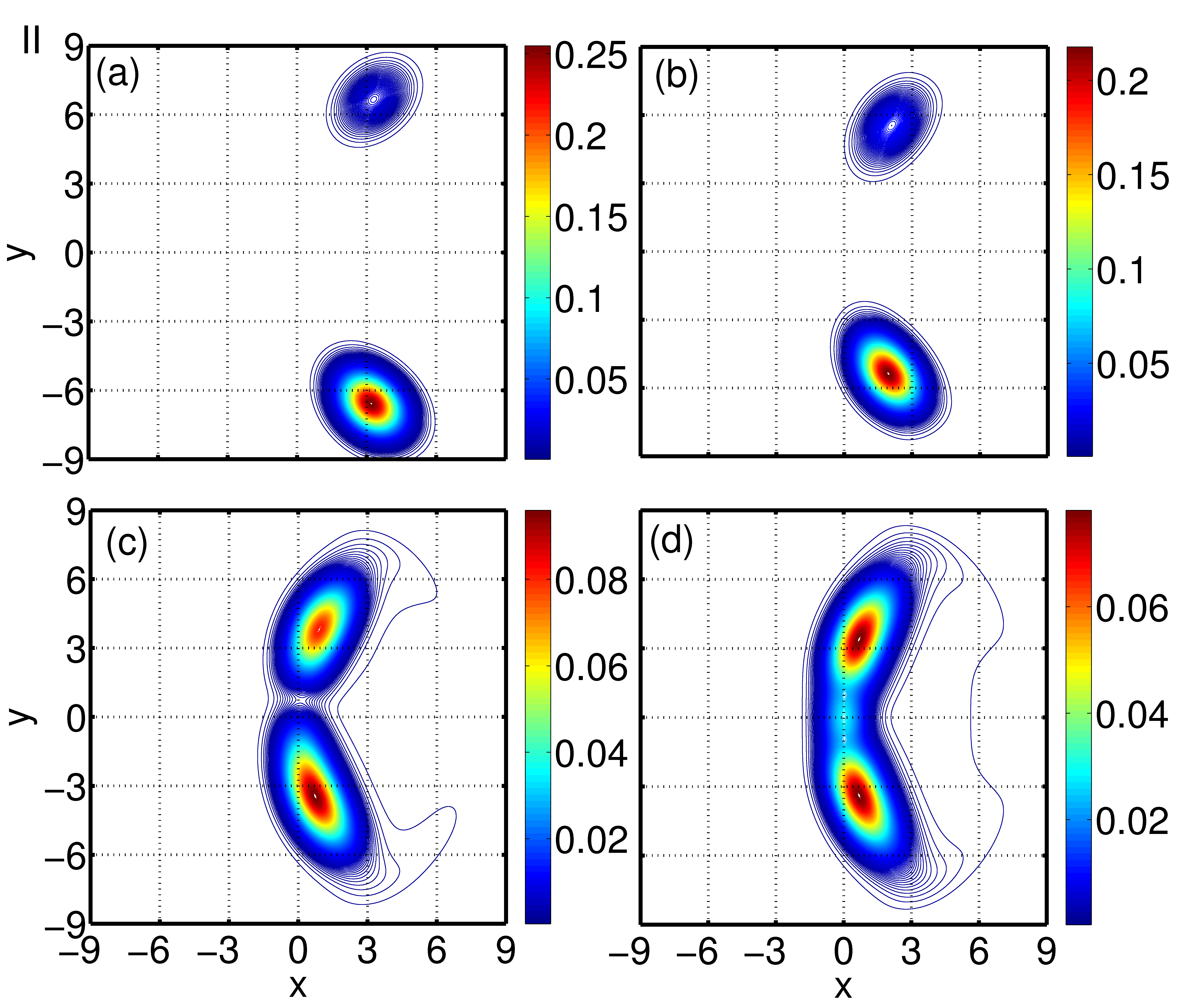}
\caption{{\it Boundary of the first-order phase transition in}  {\bf Panel I (a-c)}: Joint {\it quasi}distribution function $Q(x + iy)$ for four different points in the $(\Delta\omega_c/\kappa, \varepsilon_d/\kappa)$ phase space: \textbf{(a)} $(56.83, 16.67)$, \textbf{(b)} $(47.33, 33.33)$, \textbf{(c)} $(39.83, 50)$ and \textbf{(d)} $(0, 33.47)$. Parameters: $g/\delta=0.14$, $2\kappa/\gamma=12$ [in frames (a)-(c)] and $2\kappa/\gamma=200$ (in frame (d)), $g/\gamma=3347$. {\it Towards phase bistability at the critical point $(\Delta\omega_c=0, \,\varepsilon_d=g/2)$ in} {\bf Panel II}: Joint {\it quasi}distribution function $Q(x + iy)$ for four decreasing values of the qubit-cavity detuning to coupling strength ratio $\delta/g$: \textbf{(a)} 7.12, \textbf{(b)} 4.13, \textbf{(c)} 1.14, \textbf{(d)} 0. Parameters: $g/\gamma=3347$, $2\kappa/\gamma=200$. The driving field has a phase difference of $-\pi/2$ with respect to the drive in Fig. \ref{fig:IntAmp}(a), leading to a rotation of the distribution by that angle in the $x$-$y$ plane, as expected from  Eq. \ref{nSC}.}
\label{fig:Q1ord} 
\end{figure*}

\section{Perturbative expansion for the cavity bistability}

At first we will examine the birth of dispersive amplitude bistability for a driving frequency in the region $\Delta \omega_c \simeq g^2/\delta$ and weak drive strength. In the strongly dispersive limit the presence of the small term $g/\delta$ precludes the divergence of nonlinearity at low intracavity field amplitudes. When the length of the Bloch vector is conserved, in the absence of spontaneous emission ($\gamma=0$), the steady-state complex field amplitude is given by the relation \cite{PhotonBlockade}
\begin{equation}\label{KerrMF}
\alpha=-i \varepsilon_d \left\{\kappa - i \left[ \Delta \omega_c - \frac{g^2}{\delta} \left(1 + \frac{4g^2}{\delta^2}|\alpha|^2 \right)^{-1/2} \right] \right\}^{-1}.
\end{equation}
According to Eq. \ref{KerrMF}, we can identify $n_{\rm scale}=\delta^2/(4g^2)$ as the dispersive scale parameter. This number approaches infinity for $g \to 0$ (at constant $\delta$) and in that sense the ``thermodynamic limit'', where fluctuations vanish, is a weak-coupling limit (for a constant co-operativity parameter $C=g^2/(\kappa\gamma)$). Here, the displayed nonlinearity presents similarities to absorptive optical bistability where, setting $\gamma=0$ in the Maxwell-Bloch equation solutions {\it a posteriori}, we find the scaling parameter $\Delta \omega_c^2/(2g^2)$ \cite{PhotonBlockade} (note also that Eq. \eqref{nSC} with $\Delta\omega_c=\Delta\omega_q \equiv \Delta \omega$ is identical to Eq. 28 of \cite{PhotonBlockade}).

We apply the dispersive transformation to diagonalize the JC Hamiltonian, generating the term $\delta\sqrt{1 + n_{\rm s}/n_{\rm scale}}$, where $n_{\rm s}=a^{\dagger}a + \sigma_{+} \sigma_{-}$ is the operator of system excitations (see \cite{DispersiveTransform} and \cite{BishopJC} for more details). Expanding Eq. \eqref{KerrMF} to the lowest order in $n_{\rm s}/n_{\rm scale}$ (with $|\alpha|^2$ the semiclassical analogue of $n_{\rm s}$) yields
\begin{equation}\label{KerrApprox}
\alpha=-i \varepsilon_d \left\{\kappa - i \left[ \Delta \omega_c - \frac{g^2}{\delta} \left(1 - \frac{2g^2}{\delta^2}|\alpha|^2 \right) \right] \right\}^{-1},
\end{equation}
in agreement with Eq. (32) of \cite{PhotonBlockade}. In the dressed-cavity Duffing approximation we retain only terms up to the second-order in $n_{\rm s}/n_{\rm scale}$, and the reduced Hamiltonian acquires the quartic correction $(g^4/\delta^3)\sigma_z{a^{\dagger}}^2 a^2$ (in agreement with the semi-classical prediction of Eq. \ref{KerrApprox} for $\sigma_z=-1$). The series expansion also renormalizes progressively the driving phase space such that the effective drive strength and frequency are functions of the system operators \cite{DispersiveTransform}.

We can then derive the Wigner function for the effective dressed Duffing oscillator \cite{DuffingWalls}, calculated via the generalized $P$-representation \cite{Prepresentations, WignerKheruntsyan1, WignerKheruntsyan2}
\begin{equation}\label{WignerFinal}
W(\alpha, \alpha^{*})=\frac{2}{\pi} e^{-2|\alpha|^2} \frac{\left| _0F_1 \left(c, 2\tilde{\varepsilon}_d\, \alpha^{*} \right) \right|^2}{_0F_2(c,c^*,2| \tilde{\varepsilon}_d|^2)},
\end{equation}
where $_0F_1(\lambda;x)$ and $_0F_2(\lambda,\mu;x)$ are generalized hypergeometric functions of the variable $x$, with parameters $\lambda, \mu$. Here, $c=(\kappa - i\Delta \omega_c^{\prime}) /(i\chi)$ and $\tilde{\varepsilon}_d=\varepsilon_d/(i\chi)$ with $\chi=(g^4/\delta^3) \sigma_z$. The effective detuning $\Delta \omega_c^{\prime}=\Delta\omega_c+(g^2/\delta)\sigma_z - (g^4/\delta^3)(2\sigma_z+1)$ accounts for the correction by the dispersive shift and higher-order terms. 

We note that the perturbative expression \eqref{WignerFinal} reproduces the Gaussian form of the distribution function corresponding to a vacuum state, $W = (2/\pi) e^{-2|\alpha|^2}$, for very low driving strengths $\varepsilon_d/\kappa$, and allows us to track the progressive participation of the various nonlinear terms arising from the hypergeometric function $_0F_1$. It is therefore more instructive to write a perturbation series expansion for the numerator:
\begin{equation}\label{series}
W(\alpha, \alpha^{*})=\frac{(2/\pi) e^{-2|\alpha|^2}}{_0F_2(c,c^*,2| \tilde{\varepsilon}_d|^2)}  \left| 1 + \frac{z}{D_1} + \frac{z^2}{2! D_2}+ \cdots \right|^2,
\end{equation}
with $z=\sqrt{-8\tilde{\varepsilon}_d\, \alpha^{*}}$ and $D_m=[(c+m-1)!]/[(c-1)!]=c\cdot(c+1)\cdot \ldots \cdot (c+m-1)$, showing explicitly the development of nonlinearity for increasing drive strength. In the regime where $n_{\rm s}/n_{\rm scale} \approx 1$, the Duffing approximation breaks down as the qubit vector becomes increasingly entangled to the cavity mode, moving towards the equatorial plane in the Bloch sphere representation \cite{WallsBook}. 

In Fig. \ref{fig:WDuff}(a) we depict the cavity field distribution in the absence of entanglement with the qubit. The excitation pathways `flow' around the nodes of the Wigner function in a spiral-like fashion, as the departure from the Gaussian form becomes more apparent. These perturbative distributions approximate very well the exact ME results, in which the qubit is included as an independent degree of freedom [see Fig. \ref{fig:WDuff}(b)]. Hence, the agreement verifies the fact that the qubit participates only in dressing the cavity with a Kerr term depending on $\sigma_z=\braket{\sigma_z}=-1$. Treating $\sigma_z$ as a constant of motion when solving Hamilton's equations (for arbitrary excitation but for a short time scale in comparison to $\gamma^{-1}$) underlies the method followed by the authors of Ref. \cite{BishopJC}, who provide a semiclassical expression for $|\alpha|^2$ in the dispersive regime. The cavity response is described therein by a skewed Lorentzian curve with an amplitude dependent frequency shift, approaching the value $\sigma_z g^2/\delta$ for $|\alpha| \to 0$, and restoring the linear regime limit \cite{BlaisDisp}. Conversely, for large intracavity amplitudes and $\delta=0$, their expression yields the drive detuning corresponding to the two resonant paths, as revealed by the nonlinear equation
\begin{equation}\label{JCLorentz}
|\alpha|^2=\frac{\varepsilon_d^2}{\kappa^2 + [\Delta\omega_c \mp g/(2|\alpha|)]^2},
\end{equation}
that explains the split Lorentzian response at resonance. This distinct split has a direct relation to the origin of phase bistability, namely the formation of two quasi-independent excitation ladders with a vanishing connection between them \cite{PhotonBlockade}.   
\begin{figure}
\centering
\includegraphics[width=1.6in]{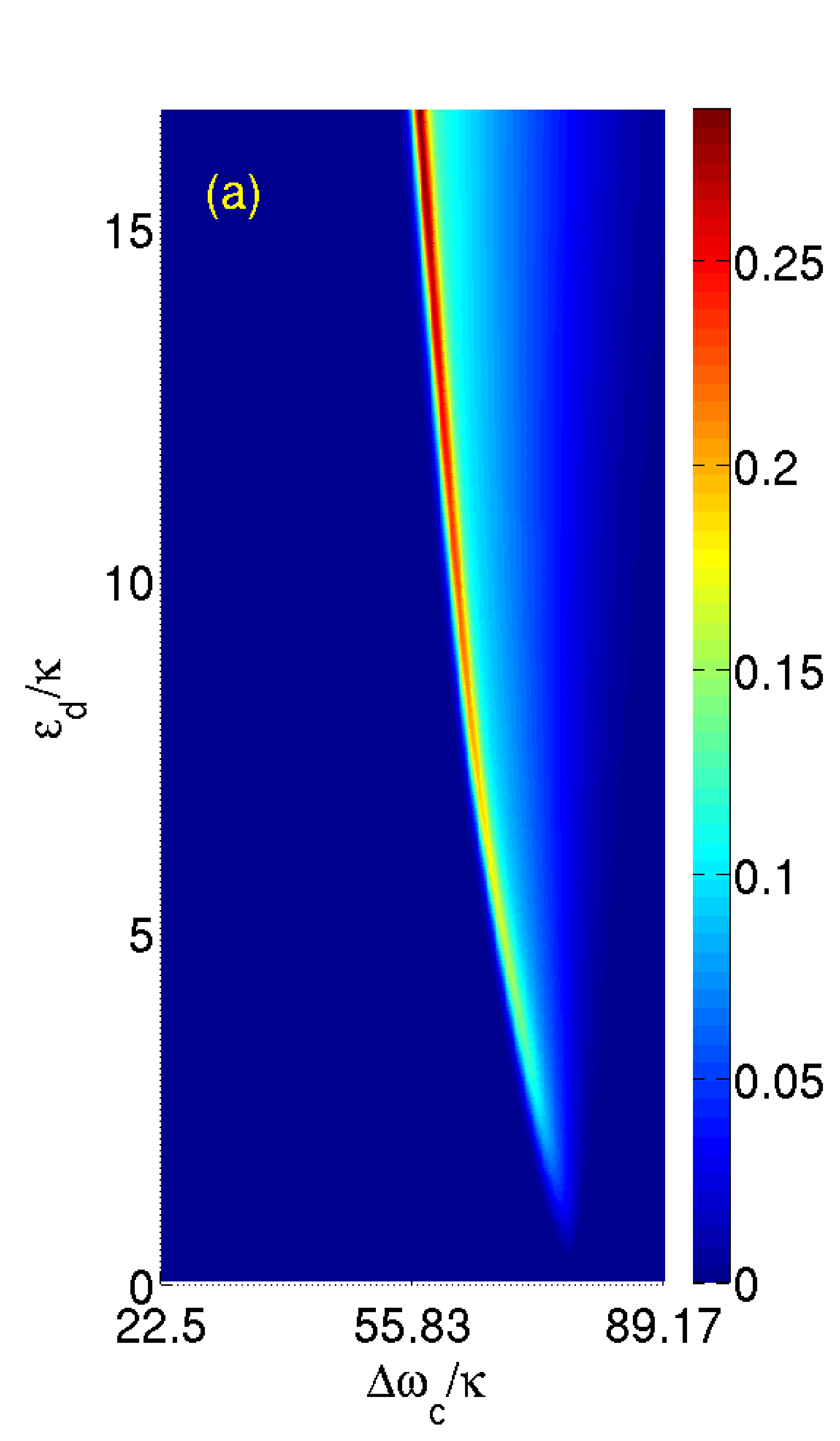}
\includegraphics[width=1.6in]{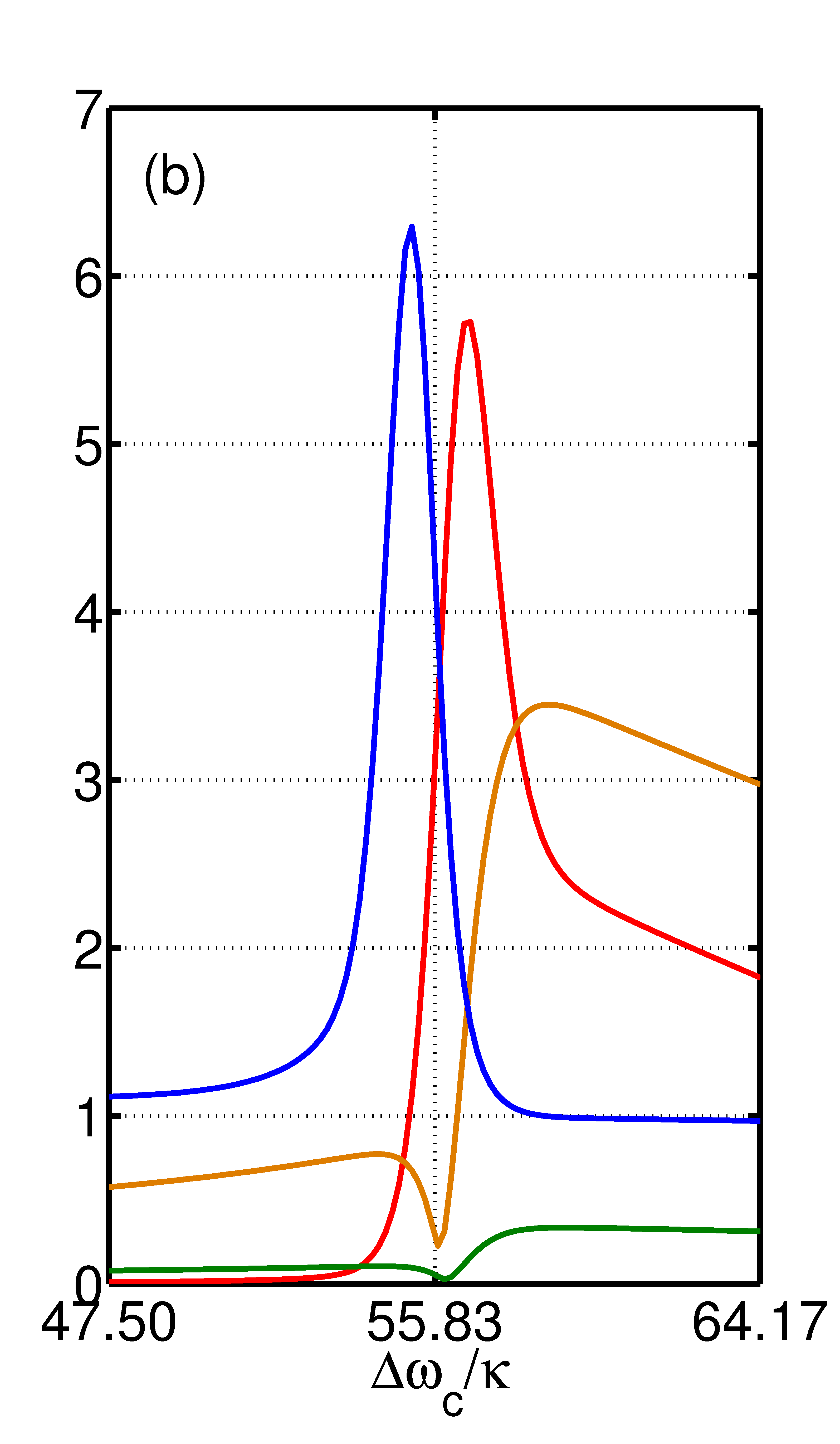}
\caption{{\it Amplitude bistability for both cavity and qubit.} \textbf{(a)} Entanglement entropy $S_q$ as a function of the driving frequency and strength. \textbf{(b)} Entanglement entropy  (x20, red curve), $\left|\braket{\sigma_{-}}\right|$ (green curve), $\left|\braket{a}\right|$ (orange curve) and autocorrelation function $g^{(2)}(0)$ (blue curve) for varying drive frequency and $\varepsilon_d/\kappa=16.67$ (corresponding to the top level of the phase space diagram in (a)). Parameters: $g/\delta=0.14$, $2\kappa/\gamma=12$, $g/\gamma=3347$, $n_{\rm scale}=12.68$.}
\label{fig:Entropy} 
\end{figure}
The region of high powers in the strongly dispersive regime can be accessed for non-demolition qubit readout with $\Delta\omega_c \sim \kappa$ \cite{Reed, BishopJC, Dispersive2}. Dynamical mapping of the qubit to the photon states has been also proposed in \cite{HighFidelity} as an optimized protocol, exploiting the first-order phase transition by means of which the photon blockade breaks down (see also the following section). 

\section{Phase transition cross-over}

We will now delineate the defining features of complex amplitude bistability with the aim of approaching phase bistability, the occurrence of which signals the ultimate difference between the effective Duffing and the full JC nonlinearity. For that purpose we plot the $Q$ function in the steady state, $Q(x + iy)=(1/\pi)\braket{x+iy|\rho_c|x+iy}$, for the reduced cavity matrix $\rho_c$ and the coherent state $\ket{x+iy}$. The region of coexisting states with probabilities of the same order of magnitude marks the boundary in the phase space of the drive where quantum fluctuations induce equiprobable transitions between the metastable states \cite{Savage, DykmanSmelyanskii}. The resulting critical slowing down is a direct consequence of nonlinear dynamics (see Chapter 5 of \cite{CarmichaelBook1}) and the departure from a Gaussian probability distribution. As the authors of Ref. \cite{Savage} note, ``bistability is a macroscopic phenomenon reached in the limit $n_{\rm scale} \to \infty$''. In our case, the bimodal distributions identify distinct states that are long-lived on the time scale $\gamma^{-1}$ (and consequently on the scale $(2\kappa)^{-1}$) even for $n_{\rm scale}=12.68$. Bimodality is depicted in Fig. \ref{fig:Q1ord} [Panel I, frames (a)-(c)], associated with maximal qubit-cavity entanglement, as we will see later on. The bright state is quadrature-squeezed along the mean-field direction (in a similar way to resonance fluorescence \cite{WallsZoller, CarmichaelBook1}), another display of the JC nonlinearity in this regime. 

In the Panel I of Fig. \ref{fig:Q1ord} we plot {\it quasi}distribution functions showing coexistent metastable states in a region where the qubit is significantly excited and approaches progressively the equator in the Bloch sphere. For increasing drive amplitude $\varepsilon_d/\kappa$ we observe a growing separation of the two metastable state distributions followed by a change in their orientation [frames (a)-(c)]. The {\it quasi}distribution function in Fig. \ref{fig:Q1ord} (d) of Panel I illustrates a precursor of phase bistability for $\varepsilon_d=g/2$, lacking nevertheless complete symmetry with respect to the horizontal axis (and hence having peaks of unequal height) because $\delta \neq 0$. We build upon this theme in the Panel II of Fig. \ref{fig:Q1ord}, where we track the emergence of phase bistability for decreasing values of $\delta/g$, and $\Delta \omega_c=0,\, \varepsilon_d=g/2$. As $\delta/g \to 0$, nonlinearity is triggered by lower photon numbers and the two peaks approach each other  (for the same values of $\varepsilon_d/\kappa$), while complete symmetry is restored only when $\delta=0$.

We proceed now to the study of entanglement as a measure of the joint participation of both quantum degrees of freedom, employing the von Neumann entropy for the reduced qubit density matrix $\rho_q={\rm Tr}_c\rho$ in the steady-state (where ${\rm Tr}_c$ denotes the partial trace over the cavity field states), defined as $S_q=-{\rm Tr}[\rho_q \ln \rho_q]=-\sum_{i=1,2}\lambda_i \ln \lambda_i$. The eigenvalues $\lambda_i$ of the reduced qubit matrix $\rho_q=(\rho_{\rm gg}, \rho_{\rm ge} \,; \rho_{\rm ge}^{*}, \rho_{\rm ee})$ are given by the expression \cite{Entanglement, Breuer}:
\begin{equation}
\lambda_{1,2}=\frac{1}{2}\left[1 \pm \sqrt{(\rho_{\rm gg} -\rho_{\rm ee})^2 + 4 |\rho_{\rm eg}|^2} \right].
\end{equation}
\begin{figure}
\centering
\includegraphics[width=3.5in]{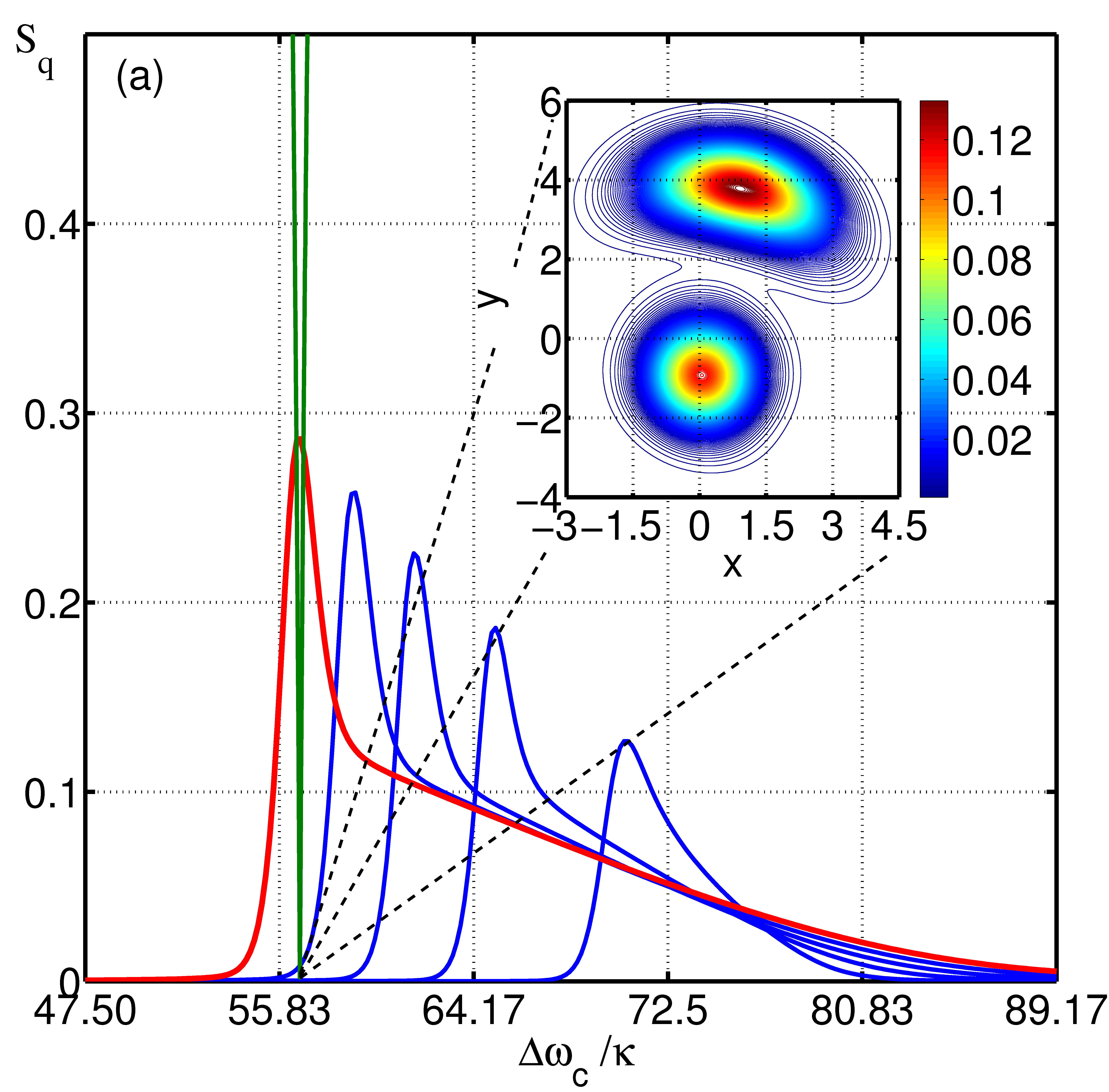}
\includegraphics[width=3.5in]{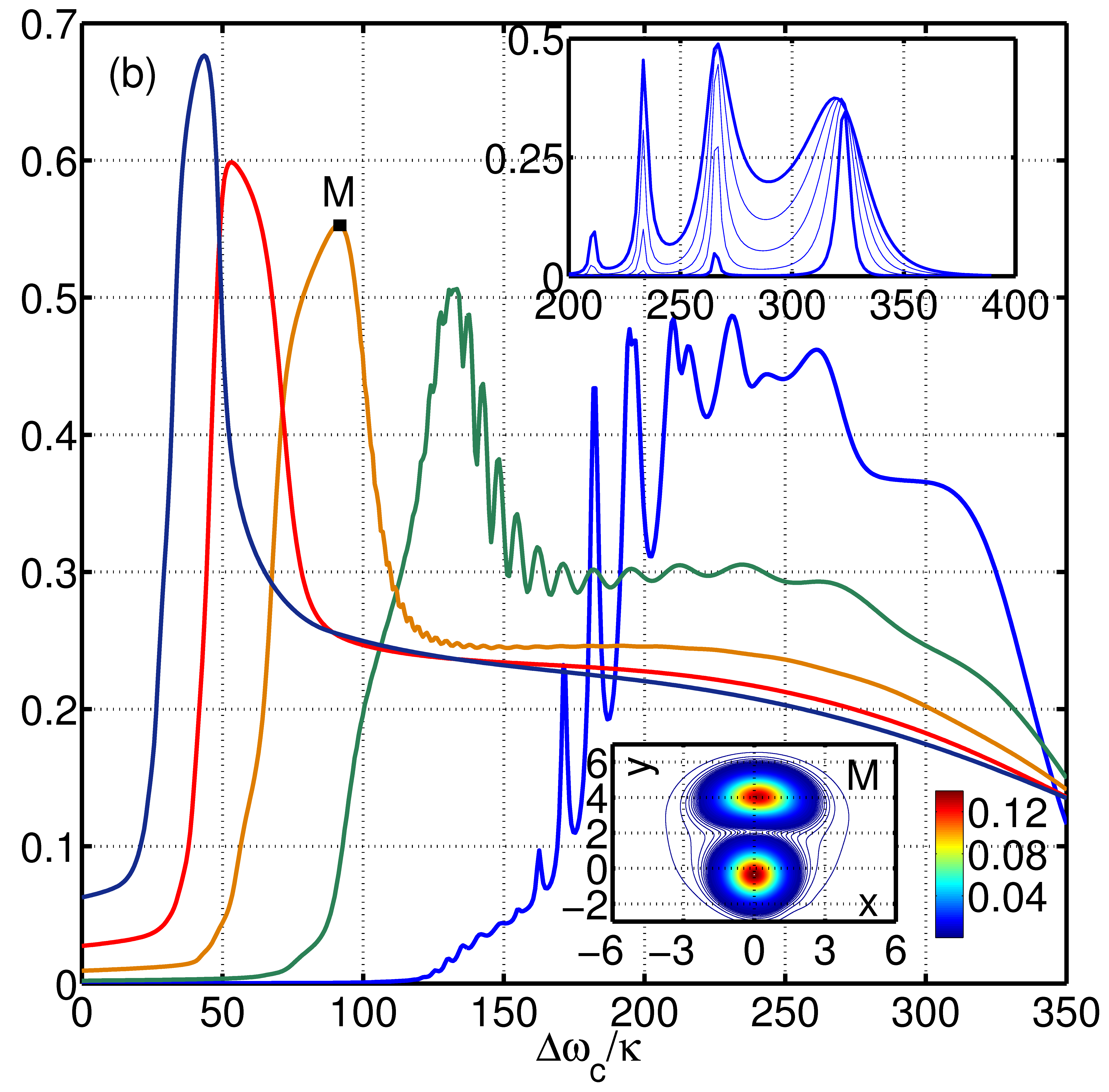}
\caption{{\it Nonlinearity for varying $n_{\rm scale}$.} \textbf{(a)} Growing entanglement entropy $S_q$ as a function of the driving frequency for $g/\delta=0.14$ (with $n_{\rm scale}=12.68$) and five equispaced drive amplitudes in the range $\varepsilon_d/\gamma=[20,100]$. The red curve corresponds to $\varepsilon_d/\gamma=100$ at the peak of which we plot the function $Q(x + iy)$ for the intracavity amplitude distribution (inset). The green curve depicts the relative difference of the $Q$ function peak values $h_1$, $h_2$ (with $h_1 > h_2$), defined as $r=(h_1-h_2)/h_1$. \textbf{(b)} Growing entanglement entropy $S_q$ as a function of the driving frequency for $g/\delta=0.87$ (with $n_{\rm scale}=0.33$) and five equispaced drive amplitudes (for each different colour, increasing in the direction: light blue, green, orange, red, deep blue) in the range $\varepsilon_d/\gamma=[200,1000]$. The upper-right inset depicts successive multi-photon resonances for the same five drive amplitudes as in (a). The bottom-right inset depicts the {\it quasi}probability function $Q(x+iy)$ corresponding to the marked point M. Parameters: $2\kappa/\gamma=12$, $g/\gamma=3347$.}
\label{fig:SqG} 
\end{figure}
The entropy $S_q$ quantifies the entanglement between the two quantum oscillators, assessing the purity of the reduced quantum state for the qubit in the steady state (following the evolution of the open system from an initial pure state). In the dispersive regime there is still appreciable entanglement between the cavity and qubit despite their strong detuning. It has recently been shown that entanglement is also present in the linear region \cite{EntDispGovia}, which we have neglected when setting $\braket{\sigma_z}=-1$ in our analytical mapping to the Duffing oscillator. The entanglement entropy tracing a first-order phase transition in the drive phase space is shown in Fig. \ref{fig:Entropy}(a). From the linear region, where entanglement is very weak [light blue region in Fig. \ref{fig:Entropy}(a) appearing at $\Delta \omega_c=g^2/\delta$], we move to the nonlinear regime where the maximum shifts to the left with a very steep drop, in a similar manner to the average photon number $\braket{n}=\braket{a^{\dagger}a}$, due to the presence of growing amplitude bistability [compare to Figs. 1 and 3(a) of \cite{PhotonBlockade}]. In Fig. \ref{fig:Entropy}(b) we plot the second-order correlation function for zero time delay, $g^{(2)}(\tau=0)$, defined via the relation $g^{(2)}(0)=\braket{n(n-1)}/(\braket{n}^2)$, in order to reveal the effect of quantum fluctuations. The peak of quantum correlations is shifted relatively to the entanglement entropy maximum, with the two curves (blue and red, respectively) intersecting closer to the position of the coherent cancellation dip in the cavity amplitude $\left|\braket{a}\right|$ [similar to the point A in Fig. \ref{fig:IntAmp}(a)] and the pseudospin projection $\left|\braket{\sigma_{-}}\right|$. The aforementioned dip has a purely quantum origin at zero temperature, which explains the amplification of quantum fluctuations in that region \cite{Bonifacio, DuffingWalls}. On the other hand, the maximum of the von Neumann entropy occurs at the frequency where the dim and bright state distributions attain peaks of equal height, as we can observe in Fig. \ref{fig:SqG}(a). When $\delta/g \to 0$ the system response becomes highly nonlinear for low drive strengths, as $n_{\rm scale}$ decreases. We observe enhanced resonant multi-photon transitions [inset of Fig. \ref{fig:SqG}(b)] gradually disappearing in the region of high drive strengths [main panel of Fig. \ref{fig:SqG}(b)]. This phenomenon is referred to as {\it breakdown of the photon blockade} [see Figs. 2(a) and 5(a) of \cite{Domokos}, and \cite{PhotonBlockade} for an extensive discussion at resonance -- $\delta=0$] accompanied by the appearance of amplitude bistability [see the $Q$ function plot in the bottom inset of Fig. \ref{fig:SqG}(b)].

\section{Phase bistability}

\begin{figure}
\centering
\includegraphics[width=3.5in]{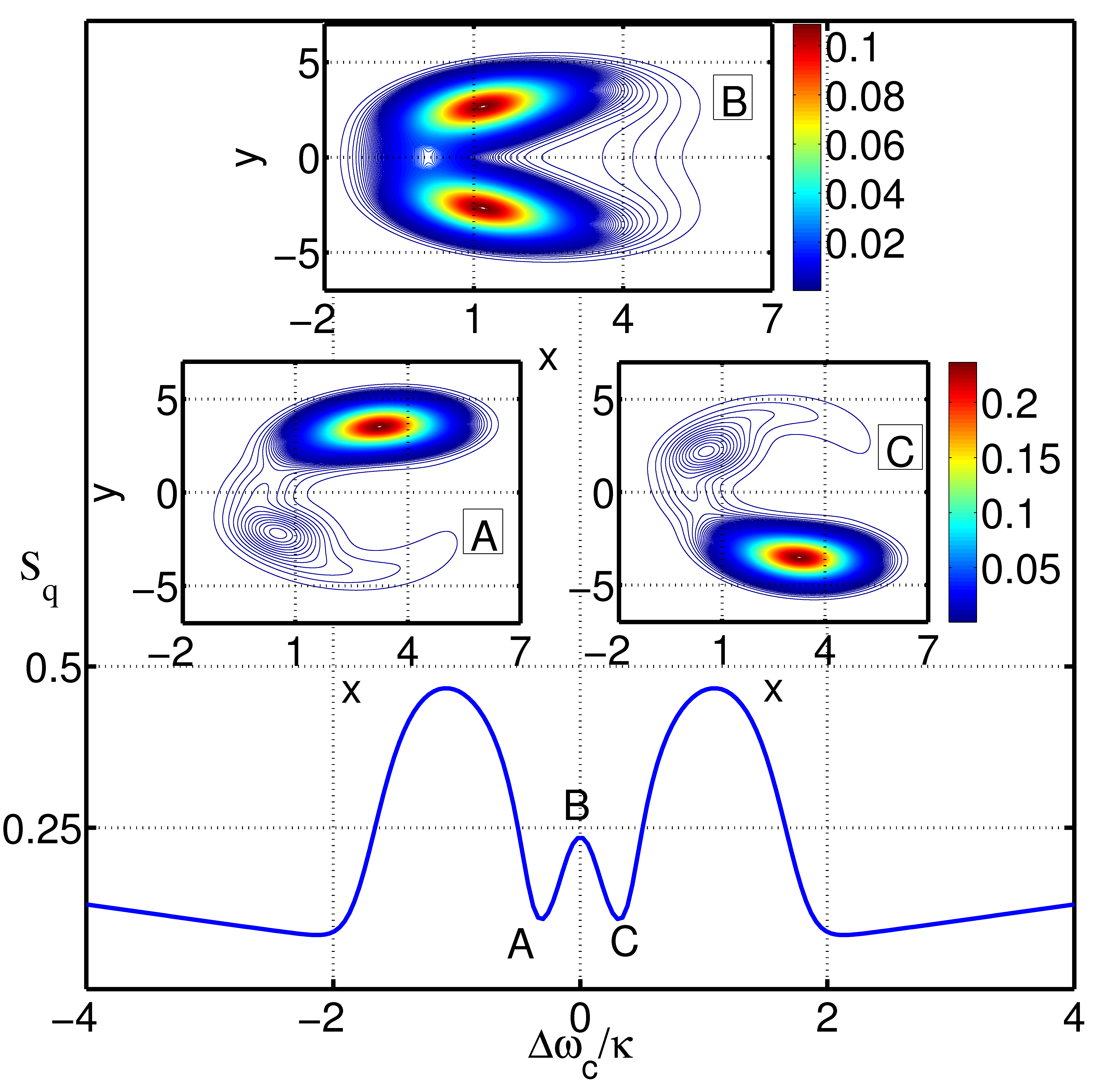}
\caption{{\it Changing excitation paths at resonance.} Entanglement entropy $S_q$ in the region of phase bistability ($\delta=0$) as a function of the driving frequency past the threshold $\varepsilon_d=g/2$ (compare with Fig. 1 of Ref. \cite{PhotonBlockade} and see Eq. \eqref{JCLorentz}, in the absence of spontaneous emission). The insets depict the joint {\it quasi}distribution function $Q(x + iy)$ for the marked points A, B and C. Parameters: $2\varepsilon_d/g=1.06$, $2\kappa/\gamma=500$, $g/\gamma=3347$.}
\label{fig:Qfinal} 
\end{figure}

Let us finally link the increasing entanglement entropy to the appearance of phase bistability past the threshold set by the critical point of the second-order quantum dissipative phase transition: ($\Delta\omega_c=\delta=0,\, \varepsilon_d=g/2$). At resonance, the nonlinearity can be triggered by low photon numbers with a different scaling parameter, associated with a strong-coupling limit \cite{PhotonBlockade}, as opposed to the strongly dispersive regime. Fig. \ref{fig:Qfinal} shows the development of a phase-bimodal distribution as we cross the line $\Delta\omega_c=0$, where the entanglement entropy has a local maximum. For growing drive strength, the entropy at point B increases and the two peaks of the $Q$ function move further apart compared to their threshold position, always remaining symmetrical with respect to the horizontal axis.

At this stage, it is instructive to invoke for a final time the solution above threshold of the so-called {\it neoclassical} equations, i.e. the semiclassical equations that conserve the length of the Bloch vector \cite{PhotonBlockade} (in the absence of spontaneous emission) which are also combined to derive the steady-state expression of Eq. \ref{KerrMF} in the dispersive regime. Neoclassical theory predicts a parity-breaking transition at resonance, according to the equation

\begin{equation}\label{PhBistMF}
\alpha=-i\varepsilon_d \left(\kappa \pm i\frac{g}{2|\alpha|} \right)^{-1},
\end{equation} 

\noindent as well as a bistable qubit vector lying on the equatorial plane ($\zeta \equiv \braket{\sigma_z}=0$) with $\nu\equiv \braket{\sigma_{-}}=\pm\alpha/(2|\alpha|)$. In that regard, phase bistability corresponds to maximally entangled states of the two coupled quantum degrees of freedom, in which the qubit polarization and the cavity field are not enslaved to the external drive, as already predicted by the mean-field analysis of Ref. \cite{SpontaneousDressedState}.\\

\section{Conclusion}

In this letter we have examined the interplay of qubit-cavity entanglement and cavity bimodality when connecting the dispersive and the resonance regimes in the driven dissipative Jaynes-Cummings model for varying qubit-cavity detuning. For the assessment of the cavity nonlinearity we have employed both the mean-field and the Master Equation treatment including quantum fluctuations. We have followed the change of the intracavity field {\it quasi}distribution functions from the strongly dispersive regime to the gates of a critical point related to a second-order quantum phase transition at resonance. We have also included in our discussion the complex amplitude bistability encountered in the driven dissipative Duffing oscillator, adopting a perturbative approach for weak driving fields. This is a region of minimal entanglement and very weak qubit involvement in the formation of the system nonlinearity, for the quantum description of which we have employed an analytical form of the Wigner {\it quasi}distribution function. The growing participation of both coupled quantum degrees of freedom marks the passage from a first-order to a second-order dissipative quantum phase transition. \\

\textbf{Acknowledgments.} The author wishes to thank H.~J.~Carmichael and E.~Ginossar for
  inspiring discussions. He acknowledges support from the Engineering and Physical Sciences Research Council (EPSRC) under grants EP/I028900/2 and EP/K003623/2.

\end{document}